\documentclass[10pt, letterpaper, reprint, superscriptaddress, aps, prl]
{revtex4-1} 

\usepackage{times}
\usepackage{graphicx}
\usepackage{siunitx}
\usepackage{float}

\DeclareSIUnit{\molar}{M}

\begin{document} 
\title{Electrokinetic Current Driven by a Viscosity Gradient}

\author{Benjamin Wiener}
 \affiliation{Physics Department, Brown University, 182 Hope St., Providence, Rhode Island 02912, USA}
\author{Derek Stein} \email{derek\_stein@brown.edu} 
 \affiliation{Physics Department, Brown University, 182 Hope St., Providence, Rhode Island 02912, USA}


\begin{abstract}

Gradients of voltage, pressure, temperature, and salinity can transport objects in micro- and nanofluidic systems by well known mechanisms.
Here we report the discovery of a transport effect driven by viscosity gradients, which cause an ionic current to flow inside a glass nanofluidic channel.
Measurements of the current are well described by a simple model wherein counterions in the electric double layers near the surfaces drift in the direction of decreasing viscosity with a drift speed equal to the gradient of the ions' local diffusivity.
Drift in a viscosity gradient is a consequence of multiplicative (state-dependent) noise, which results from a particle's thermal fluctuations depending on its position. 
This surprisingly large effect, measured in a highly controlled nanofluidic environment, reveals fundamental behavior that is relevant to a broad range of systems.

\end{abstract}

\maketitle 


Electrokinetic transport phenomena like electrophoresis, thermophoresis, and diffusiophoresis, which arise from gradients in voltage, temperature, and solute concentration, respectively, play important roles in biology, geology, and micro- and nanofluidic systems\cite{anderson1989colloid,prieve1982migration,bocquet2010nanofluidics}.
Could gradients in viscosity, which are ubiquitous in nature and technology, also drive transport?
Einstein's theory of Brownian motion showed that the viscosity of a liquid and the diffusivity of a particle within it are fundamentally related because both are manifestations of the microscopic interactions between the particle and the liquid's molecules \cite{einstein1905motion,uhlenbeck1930theory}.
One commonly models those interactions as noise, and the Brownian motion as a random walk whose step size is set by the noise magnitude. 
Thus, when a liquid's viscosity varies with position, so does the noise magnitude and the step size of a particle's random walk.
Interestingly, the physical model is incomplete until one specifies the rule for adding up the random steps.

A fundamental mathematical ambiguity, often called the It{\^o}-Stratonovich dilemma, arises in stochastic models where the noise is state-dependent (or \emph{multiplicative}). 
Depending on whether one evaluates the steps' sizes based on the noise magnitude at the beginning of each step, at the end, or somewhere in between, the particle will either drift or not drift \cite{van1983stochastic,volpe2016effective,de2012monte,lau2007state,brettschneider2011force}.
A few experiments previously concluded that the \textit{isothermal} (end-of-step) rule applies to colloidal particles, based on observations of drift in the effective viscosity gradient created by the particles' proximity to a solid surface \cite{volpe2010influence,lanccon2001drift, brettschneider2011force}; the observed drift was subtle and short-ranged because the proximity effect only stifles diffusion over distances comparable to the particle size \cite{dufresne2001brownian}.
Here, we report the discovery of \emph{viscophoresis}, an electrokinetic effect whereby a viscosity gradient generates easily measurable ionic currents inside long nanofluidic channels. 
We show the effect originates in ions subject to multiplicative noise and obeying the isothermal rule.

\begin{figure}[H]
	\centering
    	\includegraphics[width=3.4in]{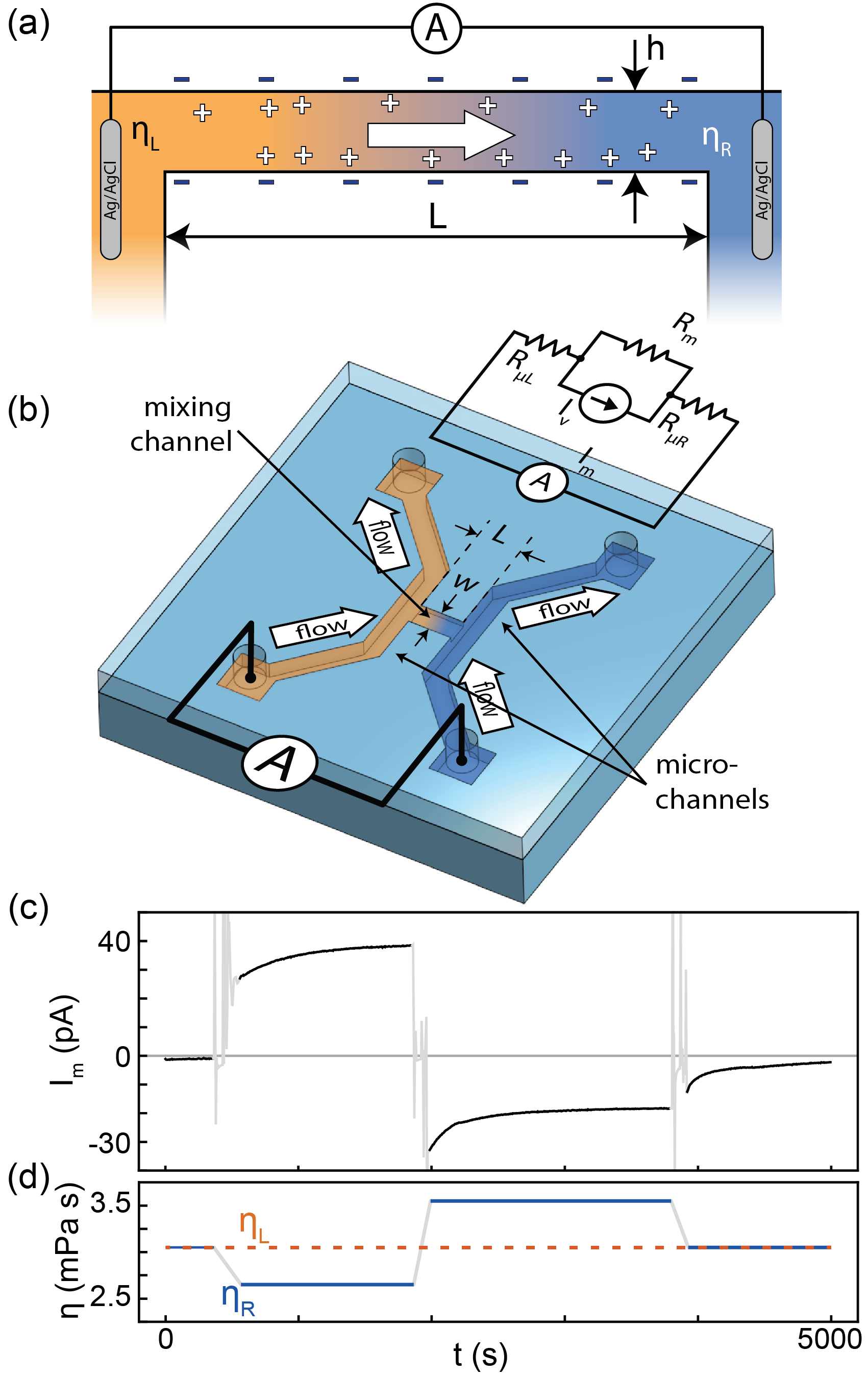}
		\caption{Measuring viscophoresis in nanofluidic channels.
        (a) Sketch of the mixing channel cross section showing counterions drifting in the direction of lower viscosity.
        (b) 3D sketch of the nanofluidic chip indicating $w$ and $L$.  
        The equivalent circuit diagram shows how $I_v$, modeled as a current source, is related to $I_m$, $R_{m}$, $R_{\mu L}$, and $R_{\mu R}$. 
        Traces of (c) $I_v$ and (d) $\eta_L$ (orange dashed) and $\eta_R$ (blue) from a typical viscophoresis measurement. 
        }
    	\label{fig:experiment}
\end{figure}

We experimentally imposed a controlled viscosity gradient in the liquid filling a glass nanochannel and measured the electrical current resulting from the drift of counterions in the electric double layers near the nanochannel's charged surfaces.
Figure \ref{fig:experiment}(a) illustrates the basic principle.
The nanofluidic device was a glass chip containing a \SI{150}{\micro\meter}-wide, \SI{50}{\nano\meter}-deep mixing channel that bridged two parallel \SI{0.5}{\micro\meter}-deep microchannels (Fig~\ref{fig:experiment}(b)). 

We fabricated devices with mixing channel lengths of $L=100$, $200$, and $\SI{400}{\micro\meter}$ using photolithography, plasma etching, and thermal wafer bonding \cite{del2009pressure}.
We used a small mixing channel height, $h=\SI{50}{\nano\meter}$, to suppress the pressure-driven flow of fluid inside.

We pumped miscible liquids with viscosities $\eta_L$ and $\eta_R$ through the left and right microchannels, respectively, which established a stable viscosity gradient inside the mixing channel. 
We used ternary liquid mixtures composed of water, formamide, and glycerol. 
The viscosity of glycerol is much higher than that of formamide (\SI{934}{\milli\pascal\second} compared with \SI{3.34}{\milli\pascal\second} at \SI{25}{\celsius} \cite{haynes2016crc}), so by varying the ratio of those liquids with the water content kept constant at 50\% by volume, we achieved viscosities ranging from 1.4 to \SI{6.1}{\milli\pascal\second}. 
We measured liquid viscosities using a ball-drop method \cite{tang2016measurements}.

\begin{figure*}
	\centering
    	\includegraphics{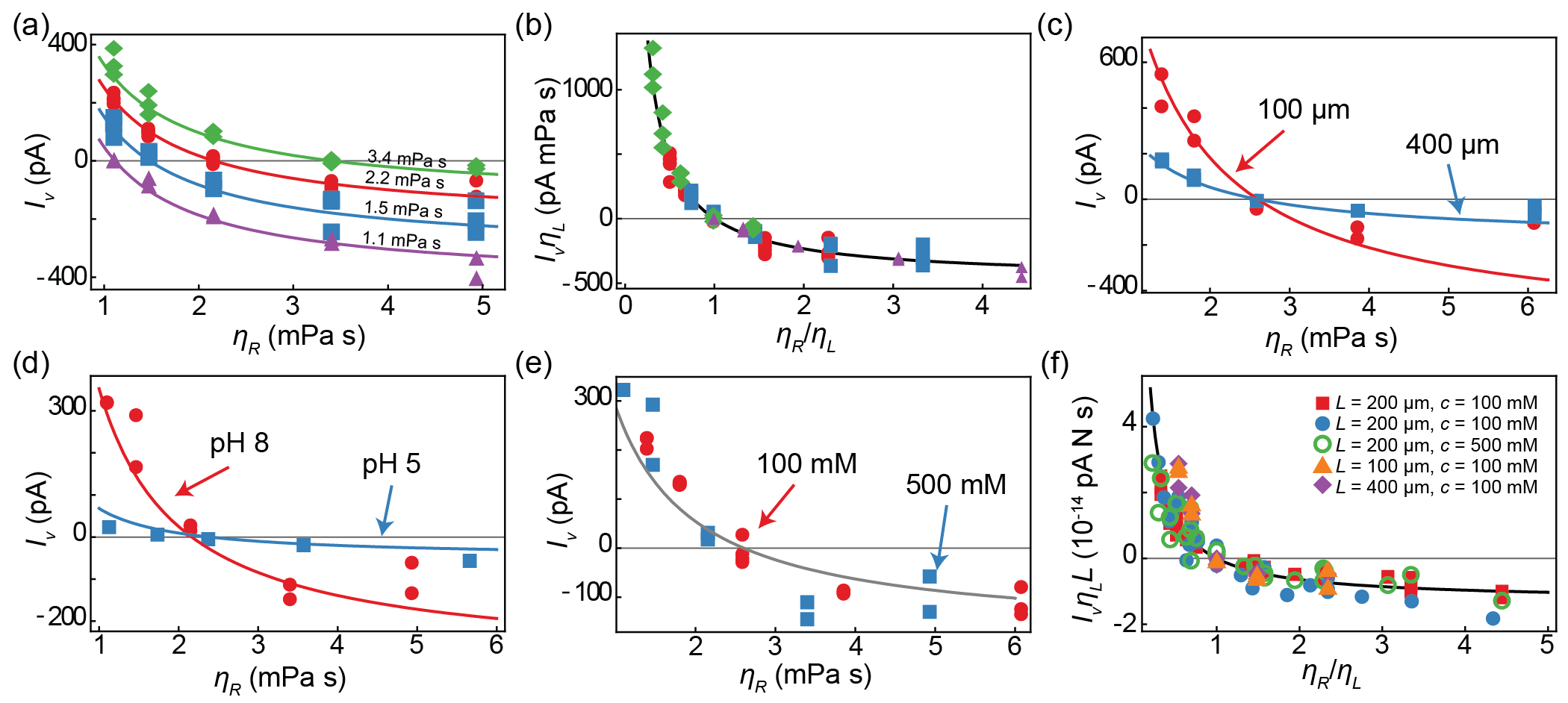}
		\caption{Experimental characteristics of viscophoresis. 
        (a) Dependence of $I_v$ on $\eta_R$ for $\eta_L=1.1$ (purple triangles), $1.5$ (blue squares), $2.2$ (red circles), and $\SI{3.4}{\milli\pascal\second}$ (green diamonds) in a $L=\SI{200}{\micro\meter}$ channel. 
        Lines show equation \ref{eqn:current} with $\sigma = \SI{200}{\milli\coulomb\per\meter\squared}$. 
        (b) Dependence of $I_v\eta_L$ on $\eta_R/\eta_L$ for the same measurements. 
        Line shows equation \ref{eqn:current} rescaled by $\eta_L$.
        The dependence of $I_v$ on $\eta_R$ is compared (c) between channels with $L=100$ (red circles) and $\SI{400}{\micro\meter}$ (blue squares); (d) between measurements at pH 8 (red circles) and pH 5 (blue squares) in a single $L=\SI{100}{\micro\meter}$ channel; and (e) between measurements at bulk KCl concentrations of $\SI{100}{\milli\molar}$ (red circles) and $\SI{500}{\milli\molar}$ (blue squares) in the same $L=\SI{100}{\micro\meter}$ channel. 
        In all cases $\eta_L=\SI{2.6}{\milli\pascal\second}$ and the lines show fits of equation \ref{eqn:current} to the data, which obtained (c) $\sigma = \SI{400}{\milli\coulomb\per\meter\squared}$, (d) $\sigma = \SI{280}{\milli\coulomb\per\meter\squared}$ for pH 8 and $\sigma = \SI{50}{\milli\coulomb\per\meter\squared}$ for pH 5, and (e) $\sigma = \SI{250}{\milli\coulomb\per\meter\squared}$. 
        (f) Dependence $I_v \eta_L L$ on $\eta_R/\eta_L$ for measurements performed at pH 8 using various $L$ and KCl concentrations, as indicated. 
        The line shows equation \ref{eqn:current} rescaled by $\eta_L L$ with $\sigma = \SI{200}{\milli\coulomb\per\meter\squared}$.
                The experimental uncertainty in these figures is comparable to the size of the symbols.}
    	\label{fig:data}
\end{figure*}

We measured the ionic current flowing through the device, $I_m$, using an ammeter (Axon Axopatch 200B) with Ag/AgCl electrodes immersed in liquid on either side.
We typically added \SI{100}{\milli\molar} KCl to the liquids to 
conduct the current generated inside the mixing channel to the ammeter.
We chose to vary only the relative amounts of glycerol and formamide because they have nearly identical solubilities for KCl \cite{seidell1952solubilities,burgess1978metal}, which prevented a chemical potential gradient from arising.
We measured the conductivities of the liquids using a Hach EC-71 conductivity meter and their pH values using a Denver Instruments UB-10 pH meter.

Figure \ref{fig:experiment}(c) plots $I_m$ and Fig.~\ref{fig:experiment}(d) plots $\eta_L$ and $\eta_R$ during a typical experiment. 
After zeroing the ammeter on a resistive dummy load, we pumped identical liquids with viscosities $\eta_L = \eta_R = \SI{2.6}{\milli\pascal\second}$ through both microchannels.
The homogeneous viscosity condition within the mixing channel resulted in a stable current close to zero. 
Next, we imposed a viscosity gradient by flushing the right microchannel with a liquid with a lower viscosity, $\eta_R=\SI{1.8}{\milli\pascal\second}$. 
A current began to flow which settled at a stable value of $I_m = \SI{38}{\pico\ampere}$ after about 25 minutes. The polarity indicated a flow of conventional (positive) current toward the right side. 
Next, we flipped the direction of the viscosity gradient by flushing the right channel with a liquid of higher viscosity $\eta_R=\SI{3.6}{\milli\pascal\second}$, and a current $I_m = \SI{-18}{\pico\ampere}$ flowed, this time toward the left channel. 
Finally, we re-established the homogeneous viscosity condition, and the flow of current halted.


The current that a viscosity gradient generates within the mixing channel, $I_v$, is related to $I_m$ by $I_v=I_m\frac{R_{\mu L}+R_{\mu R}+R_{m}}{ R_{m}}$ according to the equivalent circuit in Fig.~\ref{fig:experiment}(c), where $R_m$, $R_{\mu L}$, and $R_{\mu R}$ are the resistances of the mixing channel, left microchannel, and right microchannel, respectively. 
We observed that the conductivities of the liquids increased slowly over time, which added an experimental uncertainty of up to 12\% to $I_v$.

Figure \ref{fig:data}(a) shows the dependence of $I_v$ on $\eta_R$ for four fixed values of $\eta_L$. 
The magnitude of $I_v$ grew with the magnitude of the imposed viscosity difference. 
The current always flowed toward the lower viscosity side. 
The data from the four sets of measurements collapse onto a single curve when the product $I_v \eta_L$ is plotted against the viscosity ratio $\eta_R/\eta_L$, as shown in Fig. \ref{fig:data}(b).

Figure \ref{fig:data}(c) compares the dependence of $I_v$ on $\eta_R$ for two mixing channels of different lengths, $L= 100$ and $\SI{400}{\micro\meter}$, with $\eta_L=\SI{2.6}{\milli\pascal\second}$. 
The $\SI{400}{\micro\meter}$ mixing channel, which was four times longer than the other, produced about one fourth of the current.

Figure \ref{fig:data}(d) compares measurements performed using liquids buffered at pH 5 with liquids buffered at pH 8. 
The magnitude of $I_v$ was approximately six times lower at pH 5 than at pH 8 for all viscosity gradients tested.

Figure \ref{fig:data}(e) shows the dependence of $I_v$ on $\eta_R$ from measurements on the same device but with two different KCl concentrations, 100 and \SI{500}{\milli\molar}. 
The salt concentration had no discernible effect on $I_v$.

Figure \ref{fig:data}(f) shows data from a variety of different experimental conditions plotted on the same rescaled axes. 
It includes data taken with 100 and with \SI{500}{\milli\molar} KCl in four different devices. 
The devices had mixing channels with lengths $L=100$, $200$, and $\SI{400}{\micro\meter}$. 
The data all collapse when the product $I_v \eta_L L$ is plotted against the viscosity ratio $\eta_R/\eta_L$.

These experiments indicate that the current originates in motion of counterions in the electric double layers near the mixing channel surfaces.
Decreasing the pH of the liquid decreased the magnitude of $I_v$ because of the lower equilibrium surface charge density of the glass and the consequently lower number of counterions. 
Changing the salt concentration caused no change in $I_v$ because that bulk property of the liquid
does not significantly affect the surfaces \cite{stein2004surface}.

The ionic currents we measured are explained by the fundamental thermal (Brownian) motion of counterions in the double layer.
That motion of is determined by a huge number of interactions with molecules of the liquid whose initial conditions are generally not knowable \cite{einstein1905motion,uhlenbeck1930theory}. 
One conventionally models the aggregate effect of those interactions as random noise and the Brownian motion with a stochastic differential equation
\begin{equation}
dx = \sqrt{2D} dW_t,
\label{eqn:stochastic}
\end{equation}
where $dx$ is the displacement of a particle over the interval starting at time $t$ and ending at $t+dt$. $W_t$ is a random Wiener process whose increments, $dW_t = W_t - W_{t+dt}$, have a Gaussian distribution with mean $\left<dW_t\right> = 0$ and variance  $\left<dW_t^2\right> = dt$ \cite{volpe2016effective}. 
The diffusivity, $D$, characterizes the magnitude of a particle's thermal fluctuation and is fundamentally linked to the viscous drag on that particle; the Stokes-Einstein equation, $D=\frac{k_B T}{6\pi\eta r}$, relates $D$ to $\eta$, the thermal energy $k_B T$, and the hydrodynamic radius of the particle, $r$.
Integrating the stochastic displacements in Eq.~(\ref{eqn:stochastic}) gives the particle's trajectory. 

\begin{figure}
	\centering
    	\includegraphics{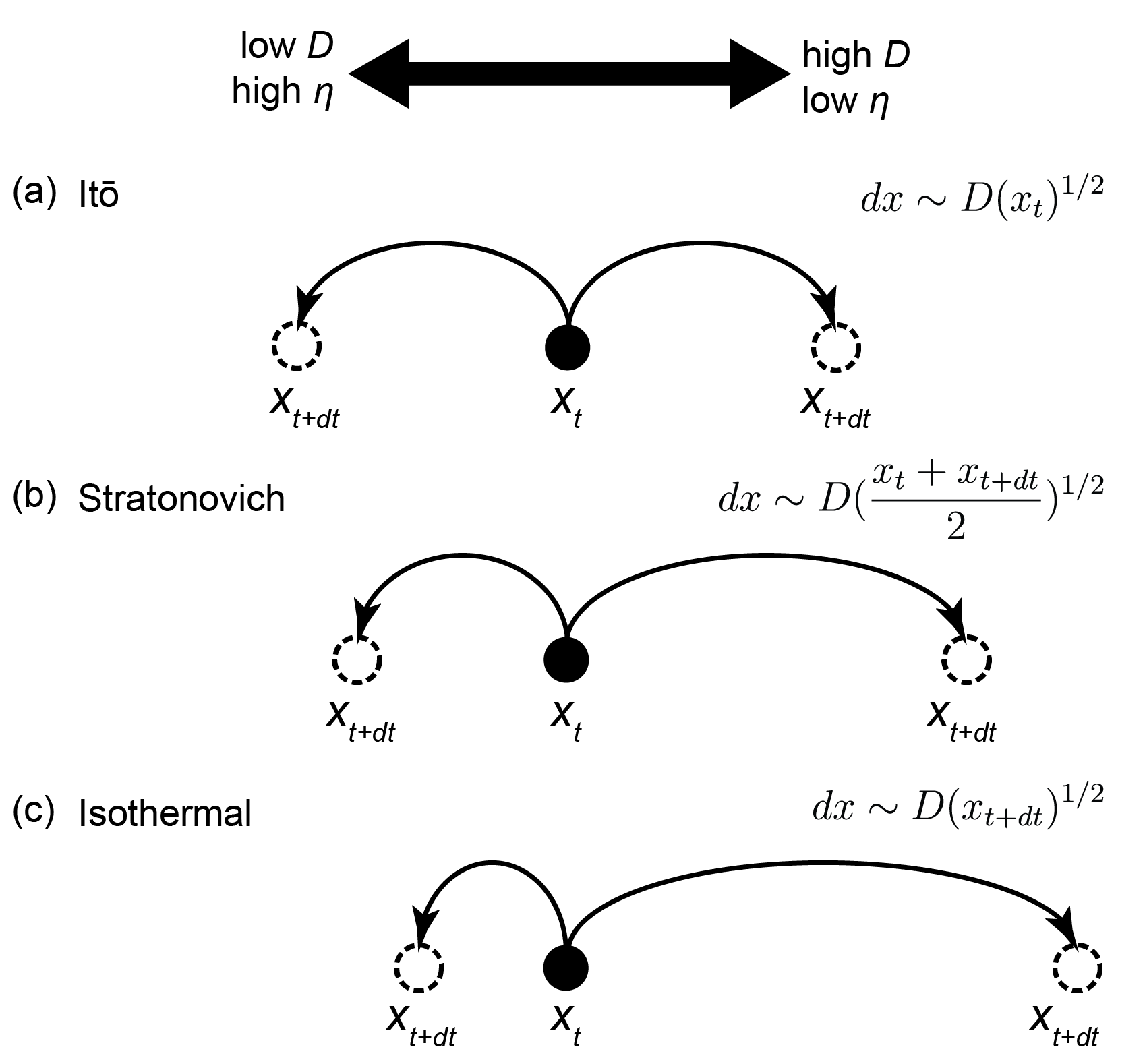}
		\caption{ Stochastic displacement models.
        Illustrations show leftward and rightward steps of random walks corresponding to (a) the It\^{o}, (b) the Stratonovich rule, and (c) the isothermal rule.}   
	\label{fig:theory}
\end{figure}

When the diffusivity varies with position, i.e. $D = D(x)$, the system is driven by \textit{multiplicative} noise -- noise whose magnitude is multiplied by a function of the system's state \cite{volpe2016effective}.
This raises a well-known mathematical difficulty called the It\^{o}-Stratonovich dilemma: Different rules for summing the stochastic displacements in Eq.~\ref{eqn:stochastic}, each one mathematically valid, result in different dynamics and therefore amount to subtly different models of Brownian motion.
In particular, one can choose to evaluate each displacement $dx$ based on $D$ at the beginning of the interval (the It\^{o} rule \cite{ito1944}), in the middle (the Stratonovich rule \cite{stratonovich1964,fisk1963quasi}), or at the end (the isothermal or H{\"a}nggi rule \cite{hanggi1982nonlinear}).
Figure~\ref{fig:theory} illustrates how these different integration rules affect a particle's dynamics.

Our measurements are well described by the isothermal rule. 
That finding is consistent with previous measurements of colloidal particles and theoretical studies of systems coupled to a heat bath \cite{volpe2016effective,volpe2010influence,lanccon2002brownian}.
A particle obeying the isothermal rule drifts toward higher diffusivity with a drift speed
\begin{equation}
\left\langle \dot{x}\right\rangle = \frac{dD(x)}{dx}.
\label{eqn:drift}
\end{equation}

The isothermal rule also leads to a generalization of Fick's law in which the flux, $J$, is related to the concentration profile, $\phi(x)$, by $J(x) = -D(x)\frac{d\phi(x)}{dx}$ \cite{de2012monte}. 
Thus, a uniform distribution of particles exhibits no net flux.
However, the electrochemical currents we measured involve the arrival and absorption of ions at one electrode and the accompanying release of ions from the other. 
That process shifts the distributions of ions away from uniformity and allows a flux in the steady state.

To provide intuition for the existence of drift and the absence of flux that the isothermal rule predicts in equilibrium, an analogy with sedimenting particles is sometimes invoked \cite{lanccon2002brownian}. 
Particles sedimenting in a container have a net drift due to gravity but reach a flux-less equilibrium supported by a concentration gradient.
However, the boundary condition for a particle arriving at the bottom of a closed container differs fundamentally from that of an ion arriving at an electrode; for a closer analogy, a sedimenting particle should be absorbed by the bottom and a new particle released from above. 
That system would clearly exhibit a steady state flux related to the drift speed and density of the particles.

We found that the currents we measured are well described by a simple relation: $I_{v} = -\sigma (2w + 2h) \left\langle \dot{x}(t) \right\rangle$, where $\left<\dot{x}\right>$ is the counterion drift speed, $w$ and $L$ are the width and length of the channel, and $\sigma$ is the average surface charge density of the channel. 
$I_v$ is related to the viscosity profile, $\eta(x)$, by eq.~\ref{eqn:drift} and the Stokes-Einstein equation. 
Assuming $w\gg h$, $I_{v} =- \sigma w\frac{k_B T}{3 \pi r \eta^2} \frac{d\eta(x)}{dx}.$
To find $\eta(x)$, we assume that the viscosity of a mixture of two liquids with viscosities $\eta_A$ and $\eta_B$ obeys $\eta(\phi_A) = \eta_B^{1-\phi_A} \eta_A^{\phi_A}$, where $\phi_A$ is the volume fraction of liquid A \cite{vignes1966diffusion}. 
We also assume that a liquid molecule's Brownian motion obeys the isothermal integration rule, which corresponds to the ``Fick'' generalization of the diffusion equation, $\frac{\partial \phi}{\partial t} = -\frac{\partial}{\partial t}D(x)\frac{\partial}{\partial x}\phi$ \cite{sokolov2010ito}. 
These assumptions lead to
$\eta(x) = \left[\frac{1}{\eta_L} - \frac{x}{L}\left(\frac{1}{\eta_L}-\frac{1}{\eta_R}\right)\right]^{-1}$ and an  expression for the viscophoretic current

\begin{equation}
I_{v} = - \frac{k_B T \sigma}{3 \pi r} \frac{w}{L} \left(\frac{1}{\eta_L}-\frac{1}{\eta_R}\right).
\label{eqn:current}
\end{equation}

Equation \ref{eqn:current} agrees quantitatively with the data in Fig~\ref{fig:data}(a) with no adjustable parameters; we measured $\sigma = \SI{200}{\milli\coulomb\per\meter\squared}$ in a separate conductance saturation experiment \cite{stein2004surface} and obtained $r=\SI{1.25}{\angstrom}$ from measurements of conductivity and viscosity using the Stokes-Einstein equation. 
Equation \ref{eqn:current} also predicts the observed $L$-dependence in Fig~\ref{fig:data}(c) using $\sigma = \SI{400}{\milli\coulomb\per\meter\squared}$ for those devices, which were fabricated together.
In Fig~\ref{fig:data}(d) we fit equation~\ref{eqn:current} to pH-dependent data using $\sigma$ as a fitting parameter and found $\sigma = \SI{280}{\milli\coulomb\per\meter\squared}$ for pH 8 and $\sigma = \SI{50}{\milli\coulomb\per\meter\squared}$ for pH 5.
As expected, the more acidic conditions lowered the surface charge density of the glass nanochannel.
Equation \ref{eqn:current} also predicts that $I_v$ is independent of KCl concentration, consistent with the data in Fig~\ref{fig:data}(e). 
The data collapse in Fig~\ref{fig:data}(b) follows directly from equation~\ref{eqn:current}, which gives $I_{v} \eta_L = - \frac{k_B T \sigma}{3 \pi r} \frac{w}{L} \left(1-\frac{\eta_L}{\eta_R}\right)$ when rescaled by $\eta_L$.
Similarly, rescaling by $\eta_L L$ gives $I_{v} \eta_L L = - \frac{k_B T \sigma}{3 \pi r} w \left(1-\frac{\eta_L}{\eta_R}\right)$, consistent with the data in Fig~\ref{fig:data}(f).

We have ruled out several alternative explanations for $I_v$.
Pressure-driven flow in the mixing channel can drive streaming currents \cite{bocquet2010nanofluidics}, but these would stop immediately upon removing the pressure; instead, $I_v$ persists for several hours after turning off the microchannel pumps. 
Furthermore, even a large \SI{100}{\milli\bar} pressure difference  across the mixing channel and relatively thin $\eta = \SI{2}{\milli\pascal\second}$ liquid would produce a negligibly small current of approximately \SI{5}{\pico\ampere}.
The Bernoulli effect can similarly generate flow and a streaming current in the mixing channel if the flow speeds in the microchannels are mismatched. 
However, this would also stop as soon as the pressure in the microchannels is relieved, and it would produce an even smaller current on the order of \SI{e-9}{\pico\ampere}.
We can exclude chemically induced currents driven by gradients in the solvation energies of the salt ions because those would increase linearly with salt concentration, whereas $I_v$ was independent of that (Fig.~\ref{fig:data}(b)).
Furthermore, solvation energy gradients should cause counterions in the double layer to flow in the direction of increasing KCl solubility, but that gives currents of the opposite polarity to $I_v$ \cite{burgess1978metal,seidell1952solubilities}.
The supplementary text presents detailed calculations of these mechanisms.






We conclude liquid viscosity gradients drove the surprisingly large ionic currents we measured inside our nanofluidic channels. 
The microscopic mechanism is the noise-induced drift of counterions obeying the isothermal rule.
The electrodes also play an essential role by permitting a finite, steady-state flux in a system that would otherwise reach a homogeneous and flux-less equilibrium.
Viscophoresis evidently uses the free energy of mixing to drive transport, in contrast with other forms of noise-driven motion, like Brownian motors, which consume chemical energy to rectify thermal noise \cite{peskin1993cellular}.
We speculate viscophoresis could cause significant motion within and between cells, across synthetic membranes, and within nanofluidic devices, where viscosities can vary by orders of magnitude over short distances \cite{kuimova2008molecular}. 
Furthermore, even viscosity gradients over large distances might influence the distributions of hydrocarbons, sediments, and other small particles in geological systems over long timescales. 
Finally, the simple picture of drifting counterions we presented describes our measurements well, despite neglecting the full behavior of co- and counterions, the electro-neutrality condition, and possibly other complications.
Computational methods can account for such details in biological, chemical, and other liquid systems where viscosity gradients naturally arise.  
It is important that they apply the isothermal rule or miss real and potentially large effects \cite{de2013translocation,lau2007state}.


We acknowledge useful discussions with Hendrick de Haan and support from Oxford Nanopore Technologies and from NSF under Grant No. 1409577. 


\bibliography{bib}
\bibliographystyle{aipauth4-1}

\end{document}